\documentclass{PoS}
\usepackage{verbatim}
\usepackage{amsmath} 
\usepackage[utf8]{inputenc}

\title{Atmospheric Charm, QCD and Neutrino Astronomy}

\ShortTitle{Atmospheric Charm, QCD and Neutrinos}

\author{Michael Benzke\\
        II.~Institute for Theoretical Physics, Hamburg University\\
        E-mail: \email{michael.benzke@desy.de}}
\author{Maria V.~Garzelli\\
        Institute for Theoretical Physics, Tuebingen University\\
        II.~Institute for Theoretical Physics, Hamburg University\\
        E-mail: \email{maria.vittoria.garzelli@desy.de}}
\author{\speaker{Bernd A.~Kniehl}\\
        II.~Institute for Theoretical Physics, Hamburg University\\
        E-mail: \email{kniehl@desy.de}}

\abstract{
We present predictions for the prompt-neutrino flux arising from the decay of charmed mesons and baryons produced by the interactions of high-energy cosmic rays in the Earth's atmosphere, making use of a QCD approach on the basis of the general-mass variable-flavor-number scheme for the description of charm hadroproduction at next-to-leading order, complemented by a consistent set of fragmentation functions. This same scheme is used for the description of charm hadroproduction at both the Tevatron and the Large Hadron Collider. We compare the theoretical predictions to those already obtained by our and other groups with different theoretical approaches. We provide comparisons with the experimental results obtained by the IceCube Collaboration 
and we discuss implications for parton distribution functions.}

\FullConference{XIII Quark Confinement and the Hadron Spectrum - Confinement2018\\
		31 July - 6 August 2018\\
		Maynooth University, Ireland}

\begin{document}

\section{Introduction}

 Neutrino Astronomy, although being a recent discipline, has already started to provide very interesting results. Very Large Volume Neutrino Telescopes (VLV$\nu$Ts) have allowed one to detect the most energetic neutrino events seen so far, in the $\mathcal{O}$(0.1--10)~PeV energy range~\cite{Aartsen:2017mau}. Even higher energies could be detected with the same instruments, with present limitations mostly driven by statistics. The advent of new VLV$\nu$Ts, like KM3NeT and Baikal-GVD, will provide precious information complementing the results already obtained by IceCube and ANTARES. Additionally, the upgrade of IceCube to IceCube-Gen2, will allow one to overcome the present statistics limitations at the highest energies and will enable more precise measurements of neutrino oscillation parameters and of further neutrino properties using high-energy atmospheric neutrinos as beams. The success of the experimental program requires parallel improvements in the modellization of both neutrino fluxes and neutrino cross sections. 

In this contribution, we focus on the case of neutrino fluxes. The IceCube Collaboration has claimed evidences for the astrophysical origin of the most energetic events they have seen so far. However, a lot of work has still to be done in order to understand the mechanisms underlying neutrino production and emission from astrophysical sources. An intriguing example is provided by the $\nu_\mu$ track event with energy $E_\nu$~$\approx$~290 TeV detected by IceCube on 22 September 2017, which was ascribed to the BL-LAC blazar TXS 0506+056, and by the follow-up of photon flairs measured by independent instruments, which have focused on the same region of the sky upon the IceCube alert~\cite{IceCube:2018dnn}. $\gamma$ rays were detected by some instruments (in particular first by Fermi-LAT and, some days later, by MAGIC), but not by other ones (e.g.\ VERITAS). The aforementioned $\nu_\mu$ event was observed by IceCube, but not confirmed by the online follow-up and time-dependent analysis of ANTARES~\cite{Albert:2018kjg}. 
Additionally, further IceCube re-analyses of old data show an enhanced $\nu$ emission from the same spatial region, already in a previous period in 2015. However, the emissions at that time were not accompanied by significant photon fluxes. The discrepancies between the observations of 2017 and those of 2015, underlying different mechanisms leading to $\nu$ emissions, represent a big challenge for the scientists expert in source study, due to the difficulty in explaining such a different behavior of a same source with time. 
On the other hand, the time integrated analysis of ANTARES of the same spatial area showed only 1 track event (actually in 2013) within an angular distance of less than 1 degree from the blazar TXS 0506+056, and no evidence for the 2015 and 2017 emissions. 
So, the situation is still unclear and the hope is to detect further high-energy events associated to possible source candidates soon, in a multimessenger approach. 

Although the study of the mechanisms underlying neutrino emissions from astrophysical sources is certainly an exciting matter, it is crucial to acquire a good control also over the background. In particular, there exists an atmospheric background due to the interaction of high-energy cosmic rays (CR) with the Earth's atmosphere. These interactions produce different kinds of hadrons,~which can decay by emitting neutrinos. 
At energies $E_\nu$ around the PeV scale, the atmospheric flux is dominated by emissions from charmed mesons. This contribution to the atmospheric flux is called ``prompt'' due to the very short time in which these mesons decay. 
In the following, we will present a QCD description of this contribution. 
We will use the general-mass variable-flavor-number scheme (GM-VFNS) approach described in Section 2 in order to compute the differential cross sections for the hadroproduction of charmed hadrons described in Section 4, which are a crucial input for the solution of the cascade equations in the atmosphere described in Section 3. Finally, in Section 5, we present our predictions for prompt-($\nu_\mu$~+~${\bar{\nu}_\mu}$) fluxes and we compare them with other predictions from the literature as well as with present limits from the IceCube experiment.

\section{GM-VFNS}

\begin{figure}[ht]
\centering
\includegraphics[width=73mm]{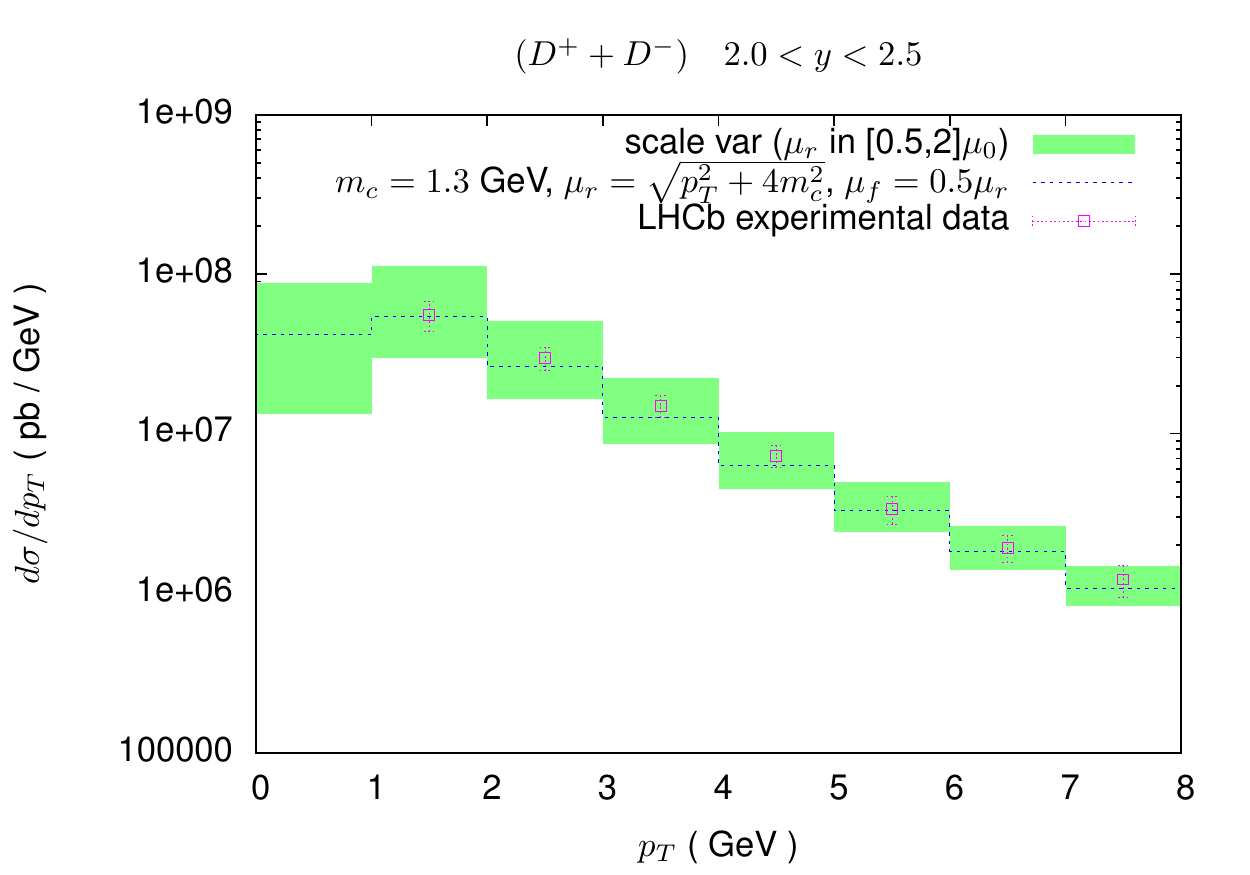}\quad\includegraphics[width=73mm]{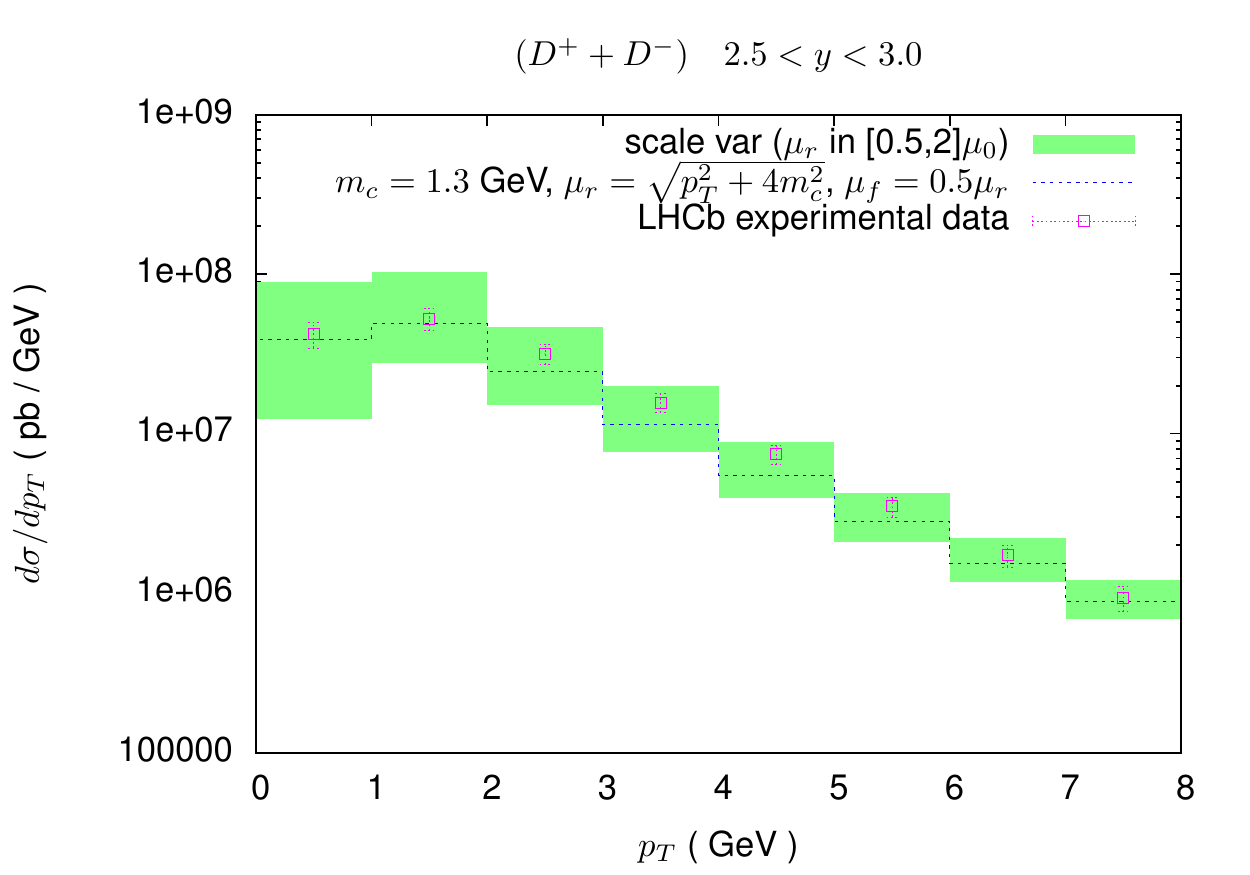}\\
\includegraphics[width=73mm]{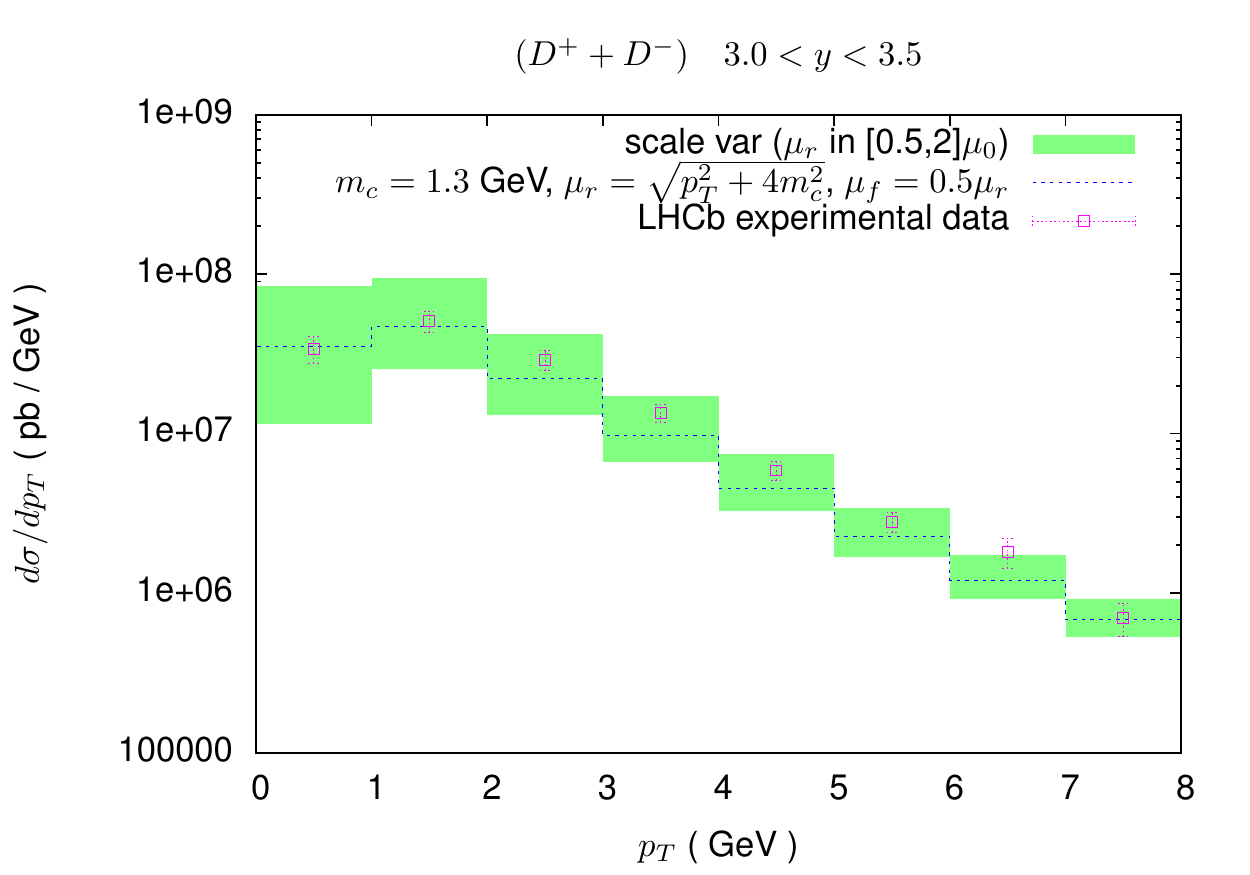}\quad\includegraphics[width=73mm]{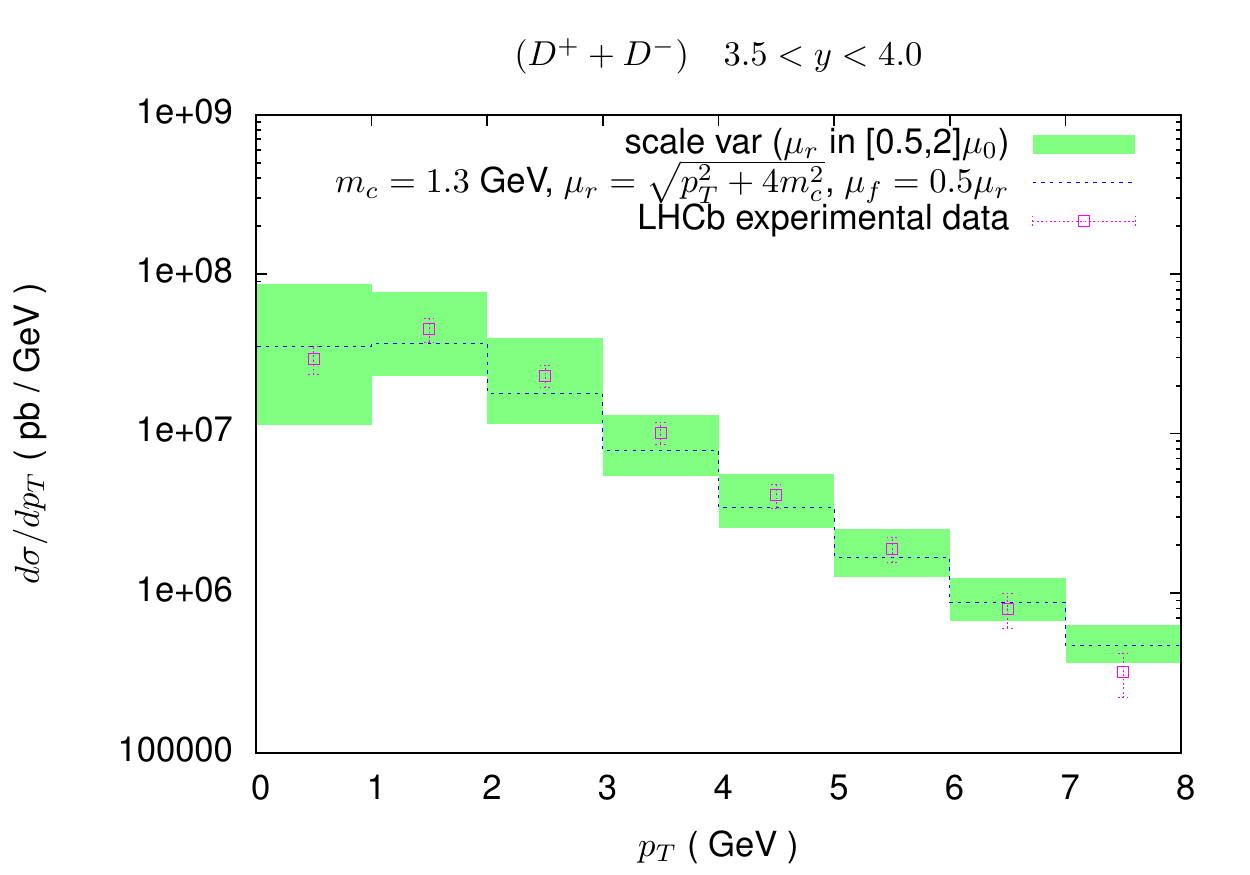}
\caption{
Our GM-VFNS predictions for the $p_T$ distributions 
of ($D^+$ + $D^-$) produced by  $pp$ collisions
 at $\sqrt{S} = 7$~TeV versus LHCb experimental data of Ref.~\cite{Aaij:2013mga}. Each panel corresponds to a different rapidity bin in the interval $2 < y < 4$. See Ref.~\cite{Benzke:2017yjn} for more details.
}
\label{fig:lhcbcompare7}
\end{figure}

When calculating cross sections of inclusive heavy-quark production, the quark mass $m_Q$ appears as a relevant scale. Depending on the kinematic region, different calculation schemes are appropriate. In the center-of-mass frame, one may introduce the produced-quark transverse momentum $p_T$ relative to the collision axis. When considering the kinematic region where $p_T$ is of the same order as $m_Q$ or lower, one uses a finite-mass or fixed-flavor-number scheme (FFNS) \cite{Nason:1987xz,Nason:1989zy,Beenakker:1988bq,Beenakker:1990maa}. In that scheme, one calculates the cross section assuming only the heavy quark to be massive while all lighter quarks are assumed to be massless and may thus appear as active flavors in the initial state. Due to the mass, there are no collinear singularities associated with the heavy quark and, consequently, no requirement to absorb them into the components of a factorized expression. Explicitly, there is no need for a fragmentation function (FF), except to model non-perturbative effects of hadronization. However, instead of collinear singularities, logarithms of the ratio of the relevant scales $\ln (m_Q/p_T)$ appear in the calculation at every order in the perturbative expansion. If one considers a kinematic region where these scales are very different from each other, the logarithms become large and may invalidate the truncation of the perturbative series at fixed order.
In the context of charm production through cosmic rays, the whole $p_T$ range is of interest in principle, and energies can become very large.
While the differential cross section in $p_T$ is dominated by the low-$p_T$ region (see e.g.\ Fig.~\ref{fig:lhcbcompare7}), at high energies, the high-$p_T$ region is still probed and may yield a noticeable contribution.

In order to make the perturbative series converge in the whole kinematic range, the potentially large logarithms can be resummed by properly factorizing the cross section and running the components to their appropriate scales. 
In
the zero-mass variable-flavor-number scheme (ZM-VFNS) \cite{Cacciari:1993mq, Kniehl:1995em, Cacciari:1995ej, Binnewies:1997gz, Kniehl:1996we, Binnewies:1997xq, Binnewies:1998vm, Kniehl:1999vf, Kniehl:2005de, Kniehl:2006mw, Cacciari:2012ny},
not only the light quarks, but also the heavy one are all considered massless and may appear in the initial state. The collinear singularities of the zero-mass calculation are absorbed into the initial-state parton distribution functions (PDFs) and the final-state FFs. Using the Dokshitzer-Gribov-Lipatov-Altarelli-Parisi (DGLAP) evolution equations, the corresponding logarithms may be resummed. However, the assumption of the heavy quark being massless is, of course, inappropriate in the low-$p_T$ region. Specifically, the calculation misses contributions proportional to $m_Q^2/p_T^2$, which are present in the FFNS approach. In summary, the differential cross section at large  $p_T$ is well described by the ZM-VFNS, while, at low and intermediate $p_T$, it becomes necessary to use the FFNS.

Both approaches may be combined using a GM-VFNS \cite{Kniehl:2004fy, Kniehl:2005mk, Kniehl:2005st, Kniehl:2005ej, Kniehl:2008zza}. Here, the terms proportional to $m_Q^2/p_T^2$ are kept in the hard-scattering cross sections, while, at the same time, the large logarithms are resummed using DGLAP evolution. The running of the PDFs and FFs is determined using the appropriate number of active flavors at each scale and performing a matching at the transition points. 

In this work, we will use the GM-VFNS as it was introduced in Ref.~\cite{Kniehl:2004fy} to compute the charm production cross sections needed to determine the prompt-neutrino fluxes. More details can be found in Ref.~\cite{Benzke:2017yjn}. The basis is formed by the factorized expression for the differential cross section of the inclusive production of a hadron $h$ in $pp$ collisions,
\begin{align}
d\sigma_{pp\to hX}(P,S) 
= F_{i/p}(x_1,\mu_i)F_{j/p}(x_2,\mu_i)\,\otimes\, d{\hat\sigma}_{ij\to kX}(p,s,\mu_r,\mu_i,\mu_f)\,\otimes\, D_{h/k}(z,\mu_f)\,,
\label{eqn:fact}
\end{align}
where $F_{i/p}$ are the PDFs, $D_{h/k}$ are the FFs, the $\otimes$ symbol denotes convolutions with respect to the scaling variables $x_1$, $x_2$, $z$, and a sum over all possible partons $i$, $j$ and $k$ is implied. 
The partonic quantities $p$ and $s$ depend on the final-state-hadron momentum $P$ and the hadronic center-of-mass energy $\sqrt{S}$ via a suitable definition of the scaling variables.
In the conventional parton model approach, the partonic cross section $d{\hat\sigma}$ is calculated assuming all partons to be massless. It will be denoted by $d{\hat\sigma}^\mathrm{ZM}$. In this case, the hadronic momenta $P_i$ are simply proportional to the partonic ones $p_i$, with the scaling variables being the corresponding factors
\begin{equation}
  p_1=x_1 P_1\,,\quad p_2=x_2 P_2\,,\quad p=P/z\,,
  \label{eqn:zdef}
\end{equation}
where $P_1$ and $P_2$ are the proton momenta, which also implies $s=x_1x_2S$.

The factorization formula still holds true in the case of non-vanishing quark masses~\cite{Collins:1998rz}. The partonic cross section $d\hat\sigma$ in Eq.~(\ref{eqn:fact}) is replaced by the corresponding finite-mass version $d\hat\sigma(m_c)$, which can be derived from the NLO parton model and the FFNS results \cite{Nason:1989zy, Beenakker:1988bq, Bojak:2001fx} in an appropriate calculation scheme.
The implementation of the GM-VFNS for hadroproduction \cite{Kniehl:2005mk} can
be presented in the following way
\begin{equation}
d\hat\sigma(m_c) = d{\hat\sigma}^\mathrm{FFNS}(m_c) - \lim_{m_c\to 0}d{\hat\sigma}^\mathrm{FFNS}(m_c) + d{\hat\sigma}^\mathrm{ZM}\,.
\label{eqn:gmvfns}
\end{equation}
The subtraction of the zero-mass limit of the FFNS result avoids a double counting with the ZM part, which contains contributions of charm quarks in the initial state. Terms proportional to $m_c^2/p_T^2$, on the other hand, are retained in the partonic cross sections.
This procedure constitutes a certain scheme choice, since the zero-mass limit of the FFNS result is not equal to the ZM one~\cite{Mele:1990cw}. This is due to the fact that the ZM calculation is performed in the $\overline{\mathrm{MS}}$ scheme, which implies a dimensional regulator $\varepsilon$, while, in the FFNS, the heavy-quark mass $m_c$ effectively regulates the collinear divergences. These two schemes do not necessarily have the same limits for $\varepsilon\to 0$ and $m_c\to 0$, respectively.
Finally, the finite-mass partonic cross sections are convoluted with PDFs and FFs as written in the factorization formula (\ref{eqn:fact}). 

Due to the form of the factorized cross section for inclusive heavy-meson hadroproduction, there appear three independent scale parameters, namely the renormalization scale $\mu_r$ and the factorization scales $\mu_i$ and $\mu_f$ corresponding to the initial and final states, respectively. A natural choice for these scales is to set them all equal to each other to $\mu_r=\mu_i=\mu_f=\sqrt{p_T^2+m_c^2}$.  
However, following this procedure leads to a badly behaved differential cross section for $p_T\to 0$. This is related to contributions with the heavy quark appearing in the initial state, calculated using the ZM-VFNS. It is, therefore, necessary to develop a method to suppress these contributions in the aforementioned limit and to retain the FFNS result, appropriately describing the cross section at small $p_T$. Recently, it has been suggested to use the freedom in the choice of the scale parameters to this end \cite{Kniehl:2015fla}. Specifically, one uses the fact that the heavy-quark PDFs vanish for a scale $\mu_i<m_c$. By setting the factorization scale for initial states to the transverse mass multiplied by a parameter $\xi_i<1$, it becomes smaller than the heavy-quark mass for small enough $p_T$:
\begin{equation}
\mu_i=\xi_i\sqrt{p_T^2+m_c^2}< m_c\quad\Leftrightarrow\quad p_T< m_c\sqrt{\frac{1}{\xi_i^2}-1}\,.
\end{equation}
In this way, the contributions with the heavy quark in the initial state are switched off for small $p_T$, and only the FFNS contributions with the heavy quark just in the final state remain.
As a result of this method, the uncertainty due to scale variations is determined by varying only the renormalization scale $\mu_r$, but keeping the initial- and final-state factorization scales fixed at their best value. 

We propose a little variant of the reasoning above. Considering that the data at low $p_T$ are better reproduced when using the $\mu_r=\sqrt{p_T^2+4m_c^2}$ functional form than the $\mu_r=\sqrt{p_T^2+m_c^2}$ one, we use the former scale definition. This was already observed in case of FFNS calculations~\cite{Garzelli:2015psa} and is related to the fact that charm quarks are always produced in pairs in the hard interaction, while in the ZM-VFNS a single charm can come out of the proton. Additionally, we fix $\mu_f = \mu_i = \mu_r/2$. For our choice of parameters, we compare differential distributions for $(D^++D^-)$ hadroproduction in different rapidity bins at $\sqrt{S} = 7$~TeV to LHCb experimental data in Fig.~\ref{fig:lhcbcompare7}. 
    
We observe that not only the LHCb data at $\sqrt{S}=7$~TeV, but also those at $\sqrt{S} = 5$ and 13~TeV are generally well reproduced when taking into account the latest revisions of Refs.~\cite{Aaij:2016jht,Aaij:2015bpa}.

\begin{figure}
\begin{center}
\includegraphics[width=0.48\textwidth]{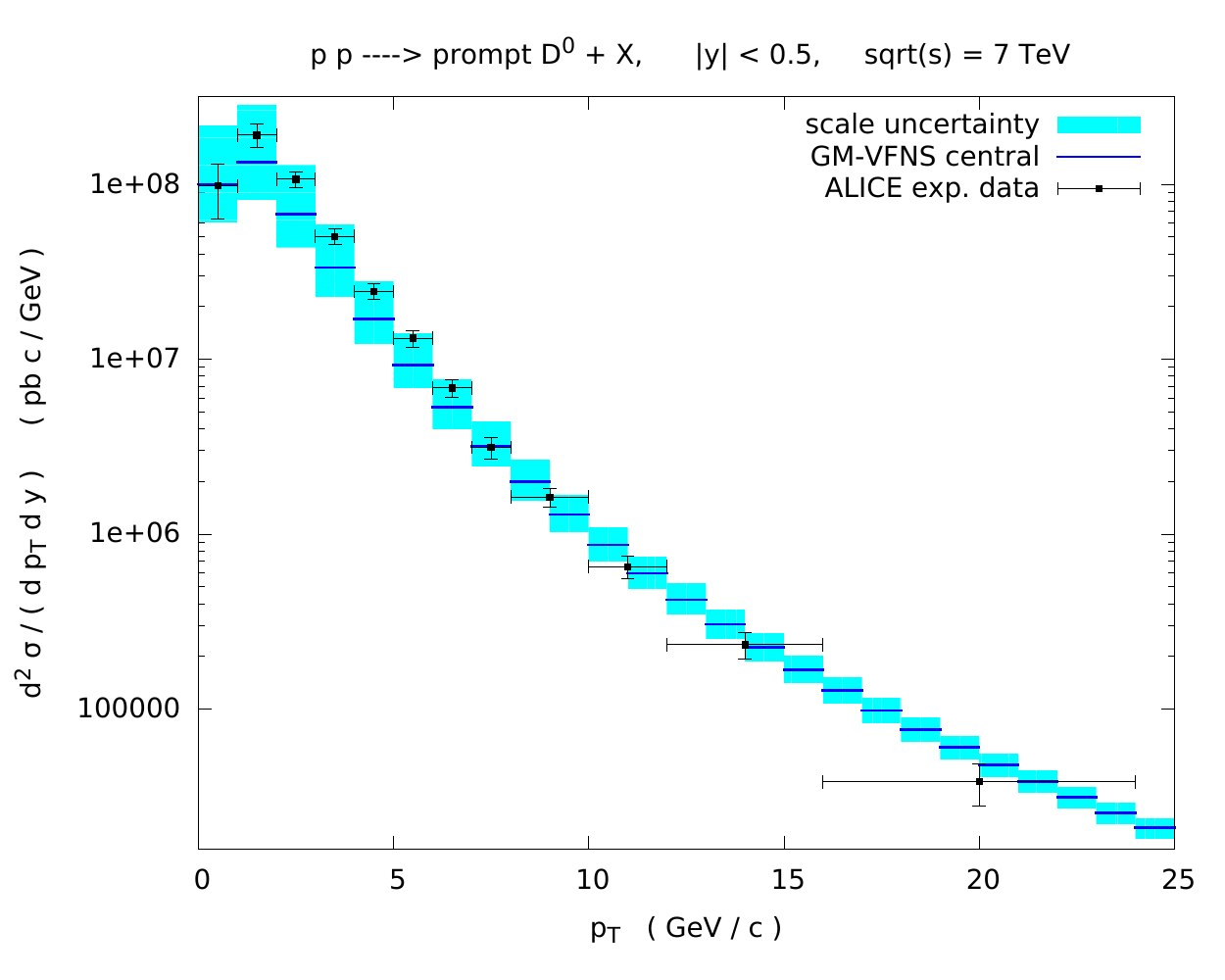}
\includegraphics[width=0.48\textwidth]{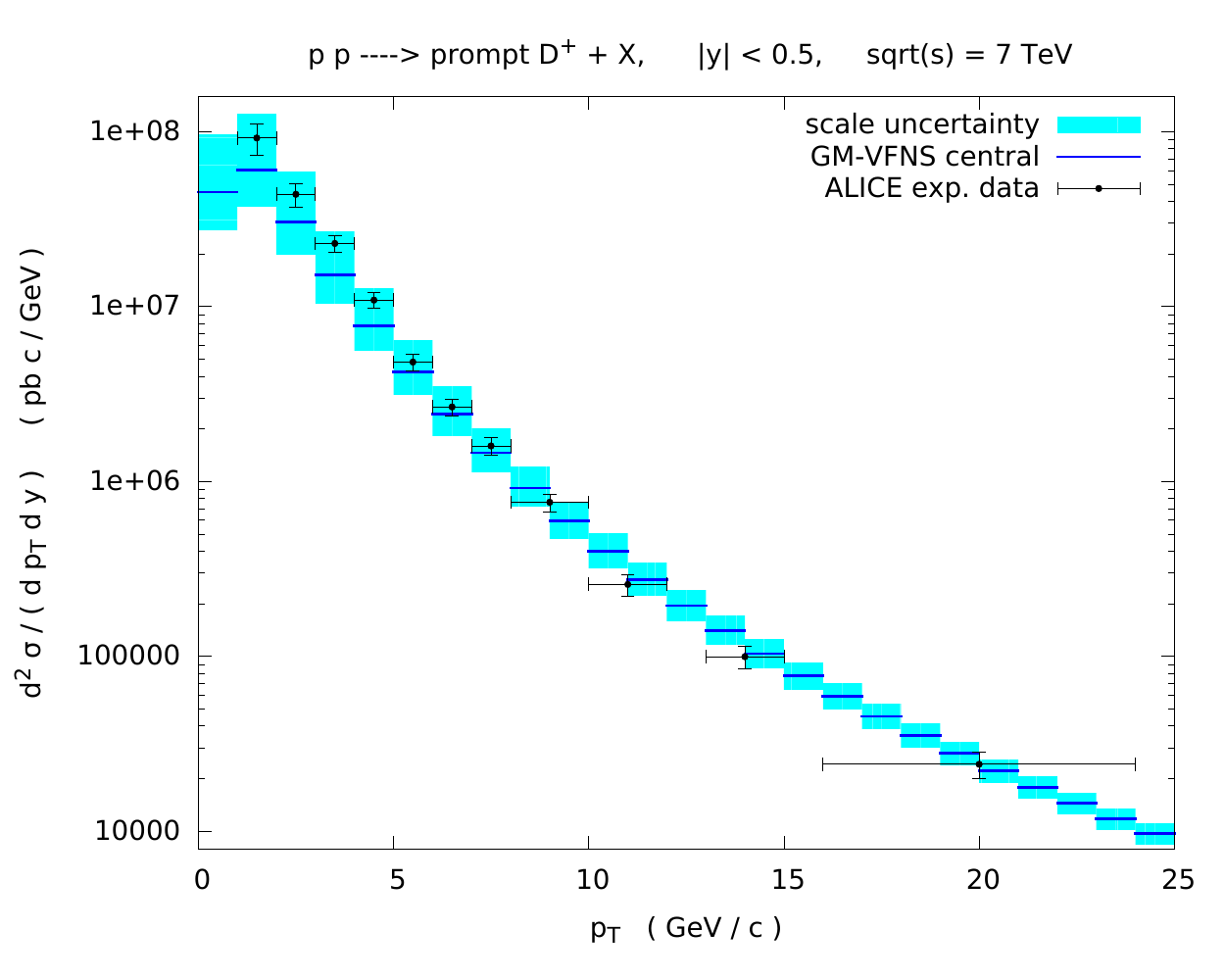}
\end{center}
\caption{\label{fig:alice}  Comparison of GM-VFNS predictions with experimental data on open $D^0$ ($\bar{D}^0$) and $D^\pm$ production in $pp$ collisions 
collected by the ALICE Collaboration at $\sqrt{S} = 7$~TeV. The theory bands refer to scale uncertainties. See the text for more details.}
\end{figure}

Using exactly the same setup adopted for the comparisons with the LHCb data, we also compare our GM-VFNS predictions with the experimental data at $\sqrt{S} = 7$~TeV released by the ALICE Collaboration, which cover a rapidity region different from the one covered by the LHCb Collaboration. We present $p_T$ distributions for $D$ mesons with $|y| < 0.5$. In Ref.~\cite{Acharya:2017jgo}, the ALICE Collaboration was able to present for the first time measurements of the $p_T$ distribution of the $D^0$ ($\bar{D}^0$) meson in the [0,1] GeV bin. We observe that the GM-VFNS predictions are in good agreement with the experimental data even in the region $p_T$ $\rightarrow$ 0, as shown in Fig.~\ref{fig:alice}.

\begin{figure}
\begin{center}
\includegraphics[width=0.48\textwidth]{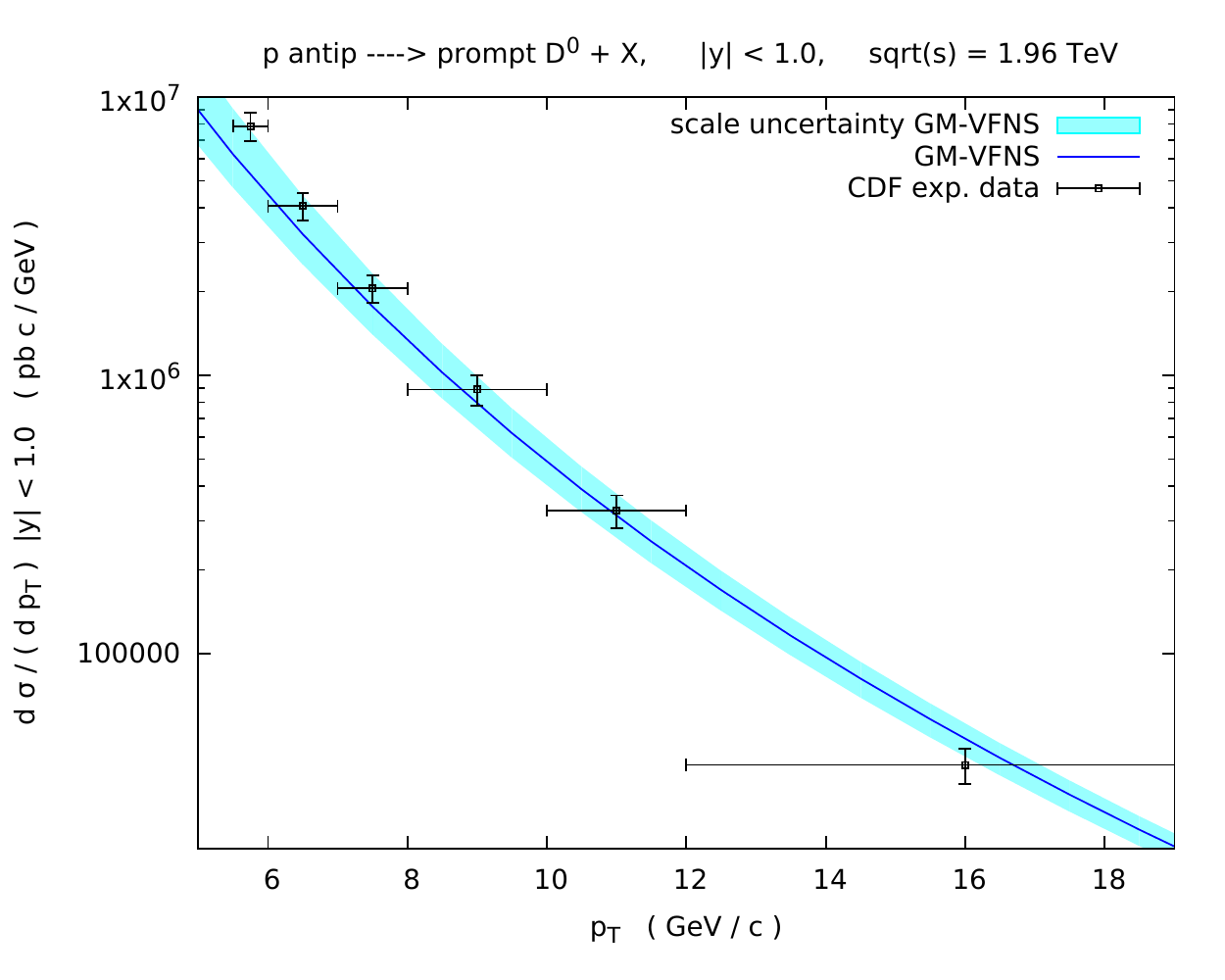}
\includegraphics[width=0.48\textwidth]{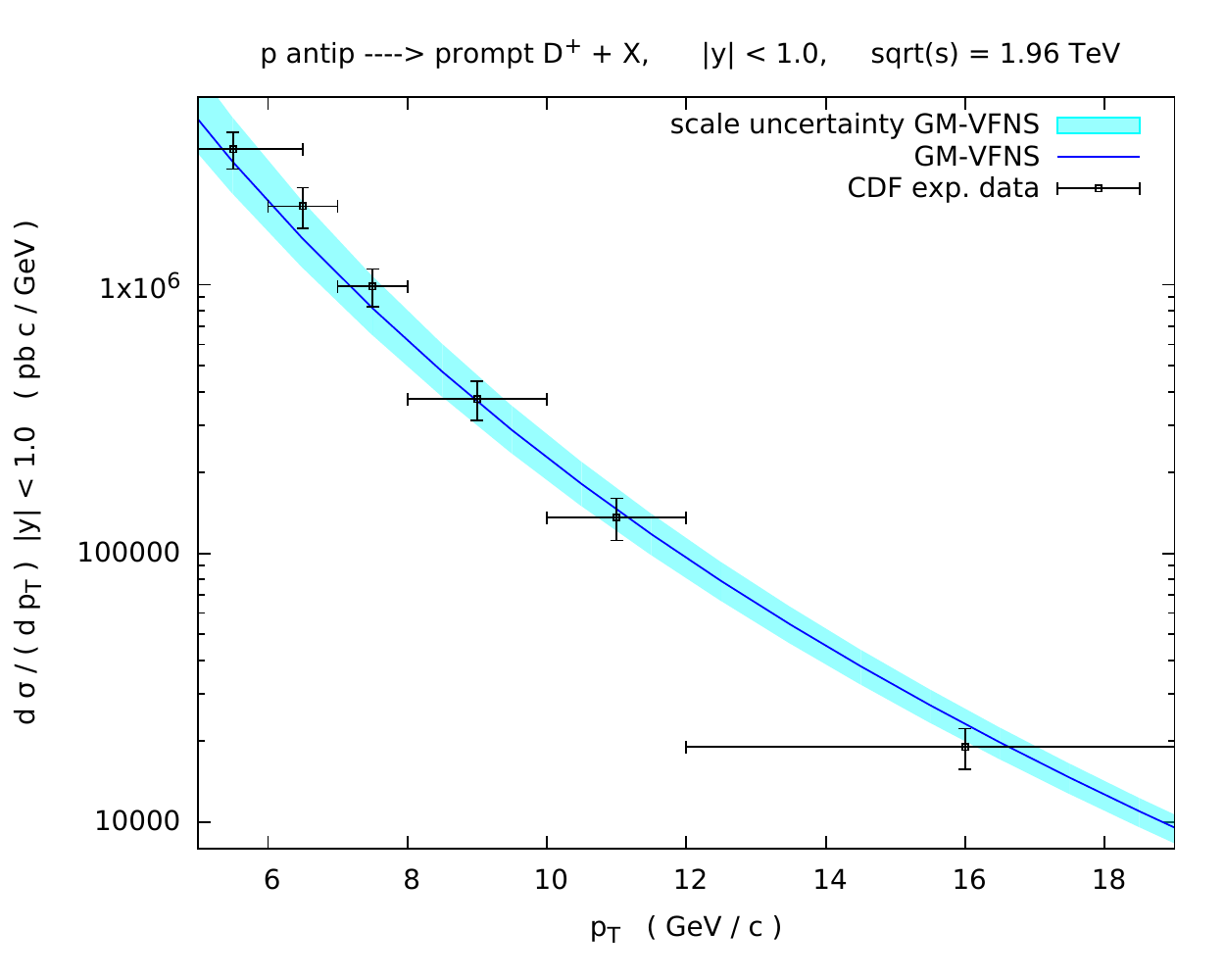}
\end{center}
\caption{\label{fig:cdf}  Comparison of GM-VFNS predictions with experimental data on open  $D^0$ ($\bar{D}^0$) and  $D^\pm$ production in $p\bar{p}$ collisions collected by the CDF Collaboration at $\sqrt{S} = 1.96$~TeV. The theory bands refer to scale uncertainties. See the text for more details.}
\end{figure}

Additionally, we compare our GM-VFNS predictions obtained using the same setup with the CDF data on $D$-meson production with $|y| < 1.0$ at $\sqrt{S} = 1.96$~TeV~\cite{Acosta:2003ax}, obtaining a very good agreement also in this case, as shown in Fig.~\ref{fig:cdf}.  
  
\section{Astrophysical Input: Cascade Equations}
A system of coupled differential equations, called cascade equations, allows us to describe the evolution of the fluxes of different particles in the atmosphere~\cite{Lipari:1993hd}: 
\begin{eqnarray}
\frac{d\phi_j(E_j,X)}{dX} & = & - \frac{\phi_j(E_j,X)}{\lambda_j^\mathrm{int}(E_j)} 
- \frac{\phi_j(E_j,X)}{\lambda_j^\mathrm{dec}(E_j)} + \nonumber\\
& + & \sum_{k\ne j} S_\mathrm{prod}^{k\rightarrow j}(E_j, X) + \sum_{k\ne j} S_\mathrm{decay}^{k\rightarrow j}(E_j, X) + S_\mathrm{reg}^{j \rightarrow j}(E_j,X) \,.
\end{eqnarray}
In these equations, $\phi_j$ is the flux of particle species $j$ at slant depth $X (l,\theta)$ = $\int_l^{+\infty} dl^\prime \rho[h(l^\prime,\theta)]$ traversed 
by the particle 
while moving from the top of the atmosphere along a trajectory with an angle $\theta$ with respect to the zenith down to a point at distance $l$ from the Earth's surface. The atmospheric profile as a function of the altitude is assumed to have an exponential form $\rho(h) = \rho_0\exp(-h/h_0)$ (isothermal model), with scale height $h_0=6.4$~km and $\rho_0=2.03\cdot10^{-3}$~g/cm$^3$. $E_j$ is the energy of the particle, $\lambda_j^\mathrm{int}$ and $\lambda_j^\mathrm{dec}$ are its interaction and decay lengths, $S_\mathrm{prod}$, $S_\mathrm{decay}$ and $S_\mathrm{reg}$ are the generation functions for its production, decay and regeneration. 

Under the assumption that the $X$ dependences of the fluxes factorize from their $E_j$ dependences, 
the generation functions can be rewritten in terms of the $Z$ moments as
\begin{align}
S_\mathrm{prod}^{k \rightarrow j} (E_j, X) &\simeq Z_{kj}^\mathrm{int}(E_j) {\phi_k(E_j, X)}/{\lambda_k^\mathrm{int}(E_j)}\,,\nonumber\\ 
S_\mathrm{decay}^{k \rightarrow j} (E_j, X) &\simeq~Z_{kj}^\mathrm{dec}(E_j) {\phi_k(E_j, X)}/{\lambda_k^\mathrm{dec}(E_j)}\,, \nonumber\\
S_\mathrm{reg}^{j \rightarrow j} (E_j, X) &\simeq~Z_{jj}^\mathrm{int}(E_j) {\phi_j(E_j, X)}/{\lambda_j^\mathrm{int}(E_j)}\,. 
\end{align}
The $Z$ moments for production and decay are defined as
\begin{eqnarray}
Z_{kj}^\mathrm{int}(E_j) & = & \int_{E_j}^{+\infty} d E^\prime_k \frac{\phi_k(E_k^\prime, 0)}{\phi_k(E_j, 0)} \frac{\lambda_k^\mathrm{int}(E_j)}{\lambda_k^\mathrm{int}(E_k^\prime)}\frac{dn(kA \rightarrow j X; E_k^\prime, E_j)}{dE_j} \, ,\label{eqn:Zint}\\
Z_{kj}^\mathrm{dec}(E_j) & = & \int_{E_j}^{+\infty} d E^\prime_k \frac{\phi_k(E_k^\prime, 0)}{\phi_k(E_j, 0)} \frac{\lambda_k^\mathrm{dec}(E_j)}{\lambda_k^\mathrm{dec}(E_k^\prime)}\frac{dn(k \rightarrow j X; E_k^\prime, E_j)}{dE_j} \,.
\end{eqnarray}
In these expressions, $dn$ is the number of particles with energies between $E_j$ and $E_j + dE_j$ produced during the interaction/decay of particle $k$ with energy $E_k^\prime$, and $A$ denotes the mass number of an air nucleus. 

Prompt-neutrino fluxes originate from the decay of heavy hadrons. In this work, we 
present the dominant contribution due to 
charmed hadrons, by considering prompt neutrinos generated by the decay of $h=h_c=D^0,\,\bar{D^0},\,D^\pm,\,D_s^\pm,\,\Lambda_c^\pm$ states, produced in $pA\rightarrow h_c+X$ scattering processes, which we approximate as a superposition of $pp\rightarrow h_c+X$ reactions. 
In the following, we focus on the ingredients for computing the $Z$ moments for the production of these hadrons. The other $Z$ moments are defined as in Ref.~\cite{Garzelli:2015psa}. 
On the one hand, the most important astrophysical ingredients 
are the CR primary spectrum and the $p$-Air total inelastic cross section, which we define as in Ref.~\cite{Garzelli:2015psa}. At present, large uncertainties affect the composition of the CR primary spectrum, especially at the highest energies. As a consequence, we use different hypotheses for the all-nucleon spectrum~\cite{Gaisser:2013bla,Thoudam:2016syr}, reflecting these uncertainties.   
On the other hand, the most important QCD ingredient is represented by differential cross sections for $D$-hadron production, as we explain in the next section.

\section{QCD Input: Hadronic Cross Sections}
\vspace{-0.35cm}
\begin{figure}[ht]
\centering
\includegraphics[width=100mm]{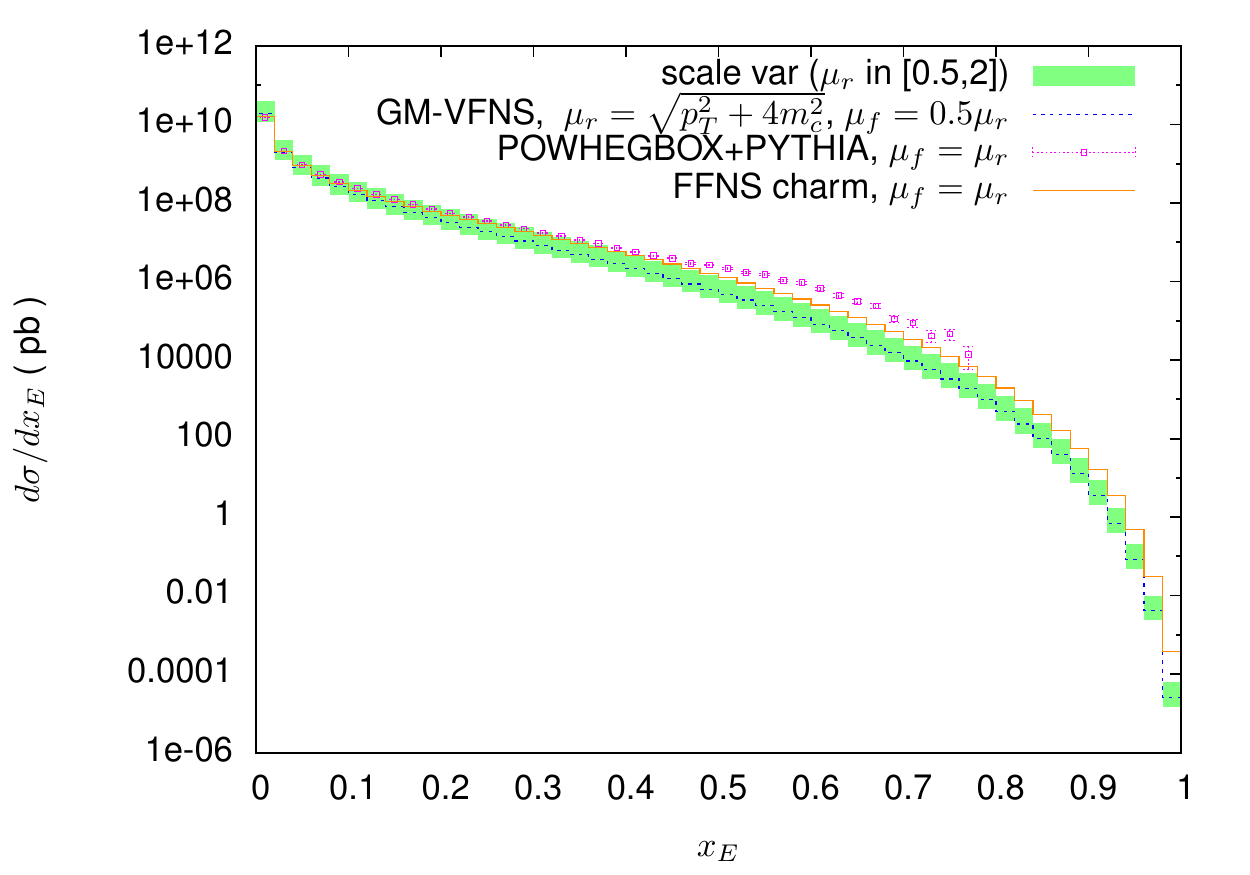}
\caption{Differential cross sections d$\sigma$/d$x_E$ for $D^0$ production in $pp$ collisions with energy $E_{p,\,\text{lab}}=10^5$~GeV in the laboratory frame
(corresponding to center-of-mass energy $\sqrt{S}\approx 433$~GeV). For the GM-VFNS prediction, we use the {\texttt{CT14nlo}} PDF, $m_c=1.3\,$GeV, and factorization scales $\mu_i=\mu_f=0.5\,\mu_r=0.5\sqrt{p_T^2+4m_c^2}$. On the other hand, the \texttt{POWHEGBOX+PYTHIA} and FFNS predictions are calculated using the natural scale choice $\mu_i=\mu_f=\mu_r=\sqrt{p_T^2+4m_c^2}$, with $p_T$ representing the charm-quark transverse momentum.
}
\label{fig:D01E5}
\end{figure}

Our predictions are based on the numerical integration of the factorization formula~(\ref{eqn:fact}) using the {\texttt{CT14nlo}} PDFs~\cite{Dulat:2015mca} and the {\texttt{KKKS08}} NLO FFs, which have been fitted at NLO to $e^+e^-$ data~\cite{Kneesch:2007ey} in the context of the GM-VFNS approach.
In the framework of high-energy physics at colliders, the cross sections are usually given as differential in the transverse momentum $p_T$ and the rapidity $y$ of the hadron evaluated in the center-of-mass frame. 
On the other hand, for the use in the cascade equations, we need to consider the laboratory frame and other differential distributions.

In Fig.~\ref{fig:D01E5}, we plot the $d\sigma/d x_E$ distribution for $D^0$ hadroproduction in $pp$ collisions at laboratory energy $E_{p,\,\text{lab}}=10^5\,$GeV, where $x_E$ is the final-state-meson to incoming-proton energy ratio in the laboratory frame. We then compare it to the central value of the same distribution obtained using \texttt{POWHEGBOX}~\cite{Alioli:2010xd}+\texttt{PYTHIA}~\cite{Sjostrand:2014zea} and to the one obtained by the FFNS approach without an FF.
We note that the predictions start to deviate for large energies, with \texttt{POWHEGBOX+PYTHIA} being the largest.
This is expected, since, if the charm quark is produced in the forward region, it can recombine with parts of the target remnant to form the charmed meson, as already observed in Ref.~\cite{Bhattacharya:2016jce}. Such an effect is not included in the factorized approach using FFs, which are fitted to $e^+e^-$ data. A Monte Carlo event ge\-ne\-ra\-tor, such as \texttt{PYTHIA}, on the other hand, implements such effects in its hadronization model \cite{Norrbin:1998bw,Norrbin:2000zc}. At small energies, there is good agreement between all predictions if one uses the standard choice $\xi_f=1$ in the \texttt{POWHEGBOX+PYTHIA} and FFNS method, while, in the GM-VFNS, $\xi_f=0.5$ allows one to appropriately adjust the low-$p_T$ behavior. 
The difference between the GM-VFNS and the FFNS approaches is due to fragmentation and the resummation of logarithms, and can be significantly reduced by use of a phenomenological FF in the FFNS calculation.

\section{Prompt-Neutrino Fluxes}

In the following, we report 
predictions for prompt-($\nu_\mu$ + $\bar{\nu}_\mu$) fluxes. 

\begin{figure}[h!]
\begin{center}
\includegraphics[width=0.62\textwidth]{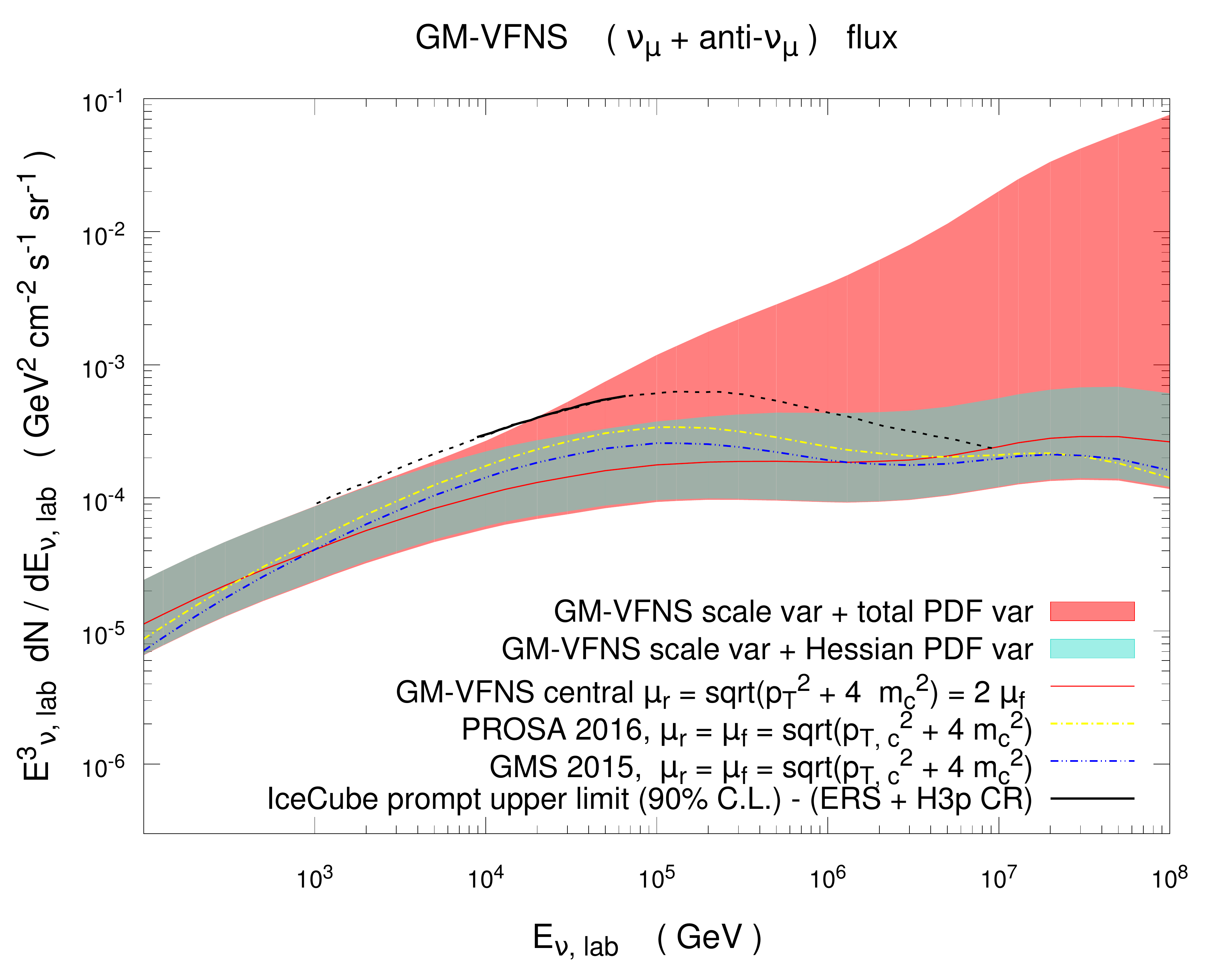}
\end{center}
\caption{\label{fig:upper}Theoretical predictions for prompt-($\nu_\mu$ + $\bar{\nu}_\mu$) fluxes evaluated in the GM-VFNS~\cite{Benzke:2017yjn} versus the IceCube upper limit~\cite{Aartsen:2016xlq}.
The theory uncertainty bands refer to scale + PDF uncertainties. The latter are computed considering all 56 member sets of the {\texttt{CT14nlo}} PDF fit, by distinguishing the Hessian members (1--52) from the non-Hessian ones (53--56). }
\end{figure}

First, we compare our predictions to the IceCube upper limit on prompt-neutrino fluxes in Fig.~\ref{fig:upper}. Uncertainty bands due to scale variation and PDF uncertainties, evaluated considering the 56 members of the {\texttt{CT14nlo}} PDF set are also shown, distinguishing the case of Hessian members 1--52 from the case of members 53--56.  
We observe that the IceCube upper limit gives indication that the {\texttt{CT14nlo}} gluon PDF uncertainties at low $x$ values (see PDF error sets 53--56), determining the behavior of prompt-neutrino fluxes at large energies and making the uncertainty band particularly large, are too large. This example points towards the possibility of using data of astrophysical origin to constrain PDF sets.

\begin{figure}
\begin{center}
\includegraphics[width=0.48\textwidth]{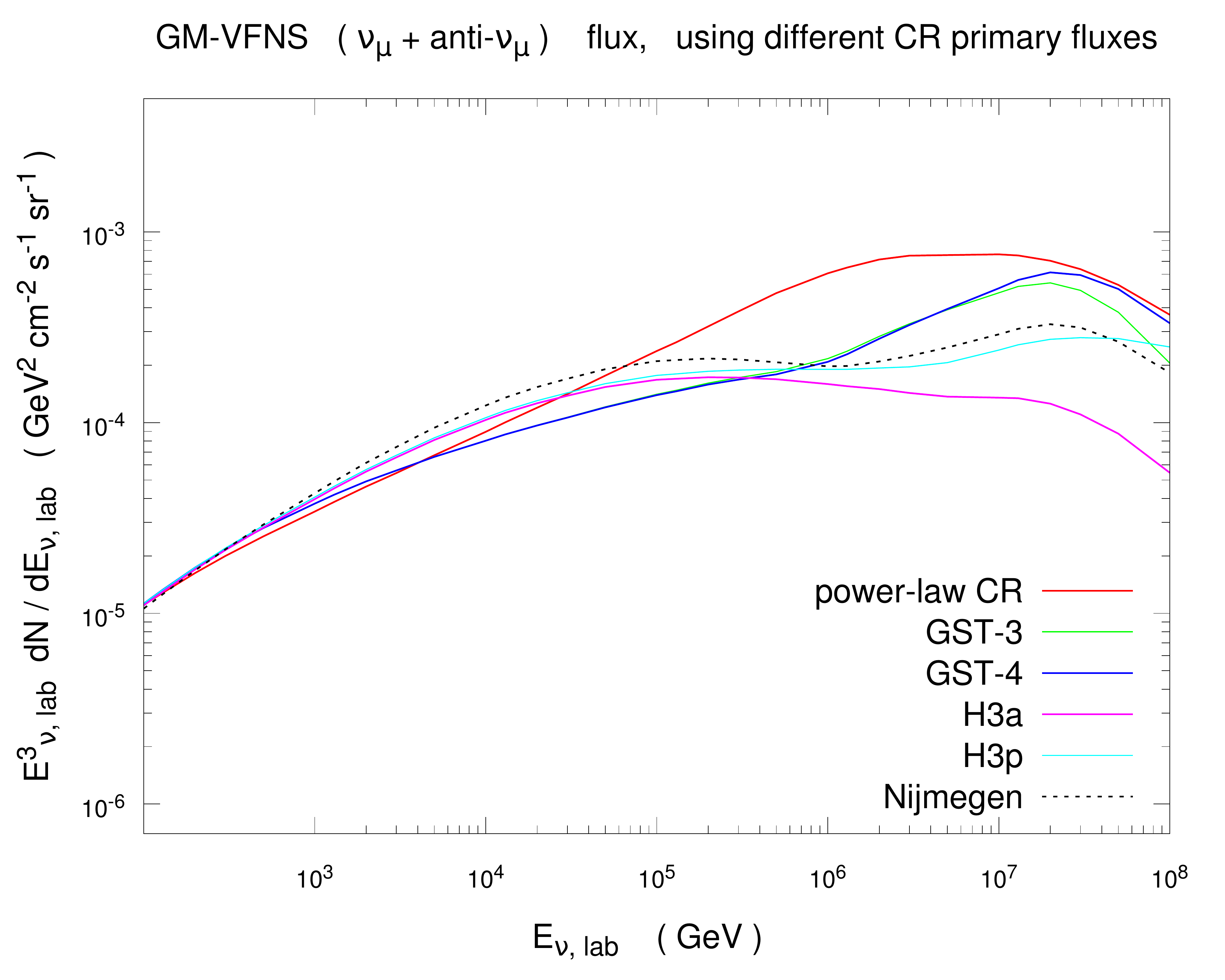}
\includegraphics[width=0.48\textwidth]{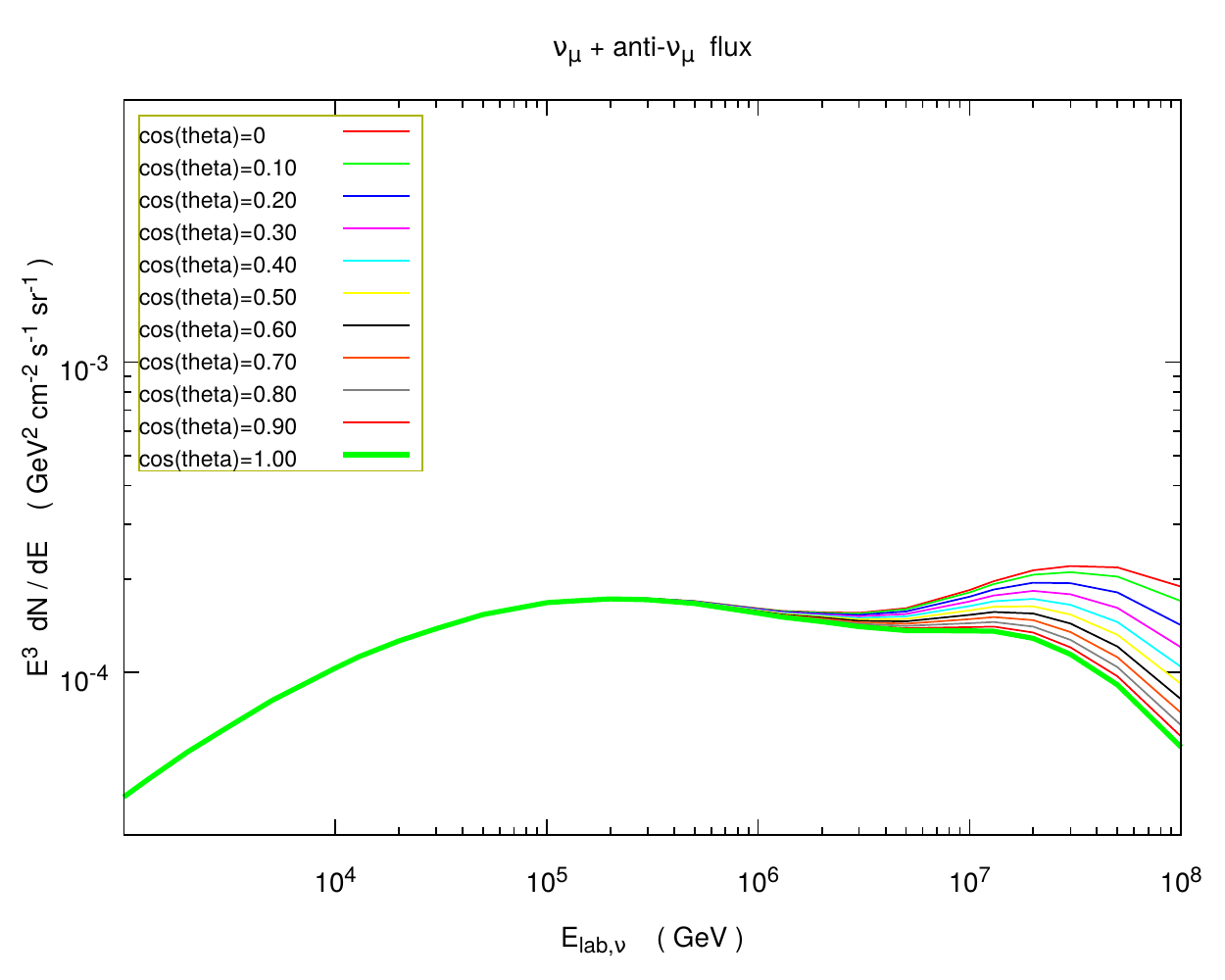}
\end{center}
\caption{\label{neufluxes} Prompt-($\nu_\mu$ + $\bar{\nu}_\mu$) fluxes, computed according the GM-VFNS approach described in this work, considering different hypotheses for the composition of the CR primary spectrum~\cite{Gaisser:2013bla, 
Thoudam:2016syr} (left panel). The dependence on the angular direction (cosine of the zenith angle) is shown for the specific case of fluxes generated by the H3a CR primary all-nucleon spectrum (right).} 
\end{figure}

Comparisons of GM-VFNS predictions obtained using different hypotheses for the CR primary spectrum are shown in the left panel of Fig.~\ref{neufluxes}. It is evident that at large energies, not only the normalization, but also the shape of prompt-neutrino fluxes depend on the assumptions regarding the CR composition, with heavy compositions giving rise to smaller fluxes than the lighter ones. Albeit to a lesser extent, at high energies, the flux depends also on the angular direction with respect to the zenith, with larger fluxes corresponding to more horizontal directions, as shown in the right panel of Fig.~\ref{neufluxes}. 

\begin{figure}
\begin{center}
\includegraphics[width=0.48\textwidth]{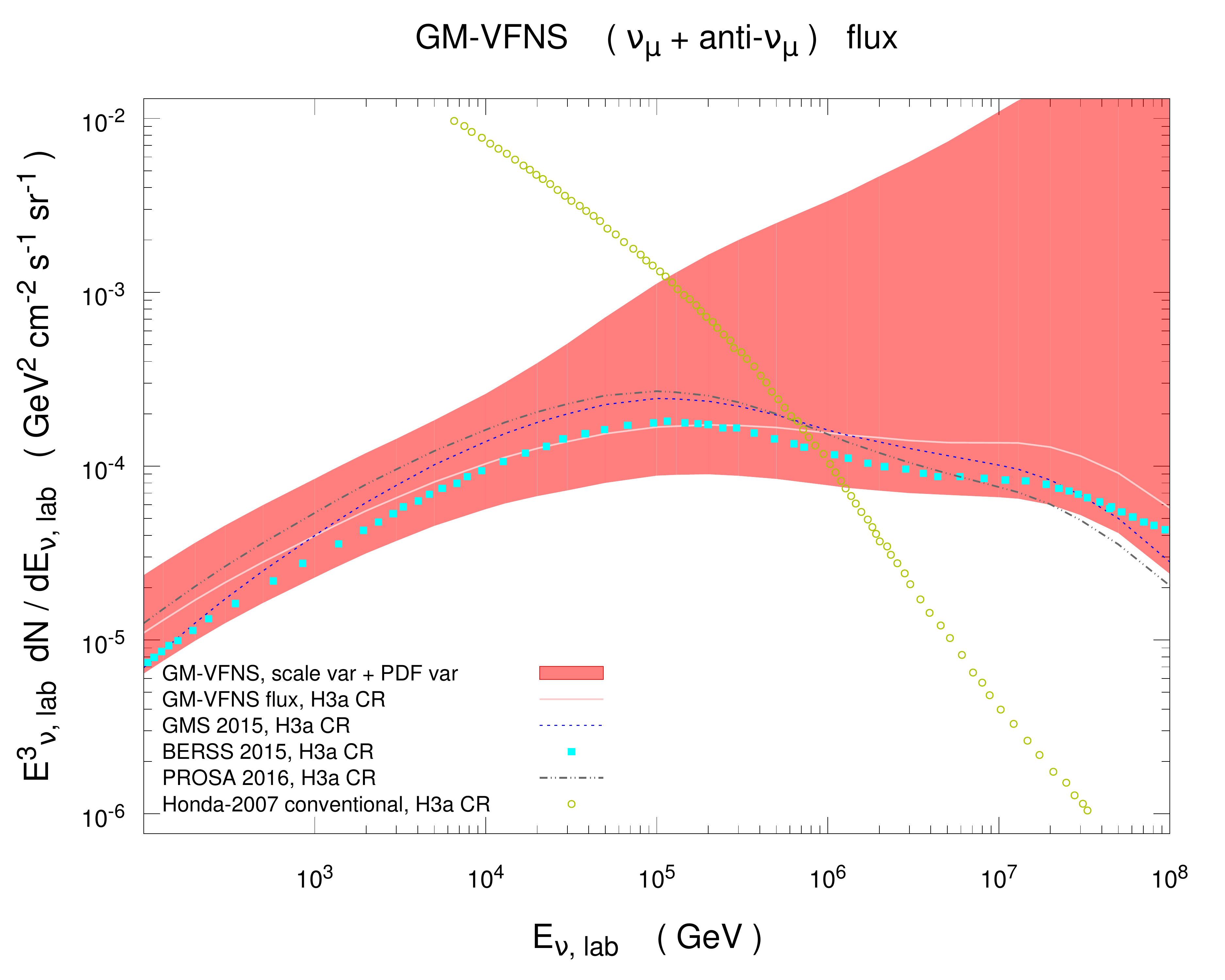}
\includegraphics[width=0.48\textwidth]{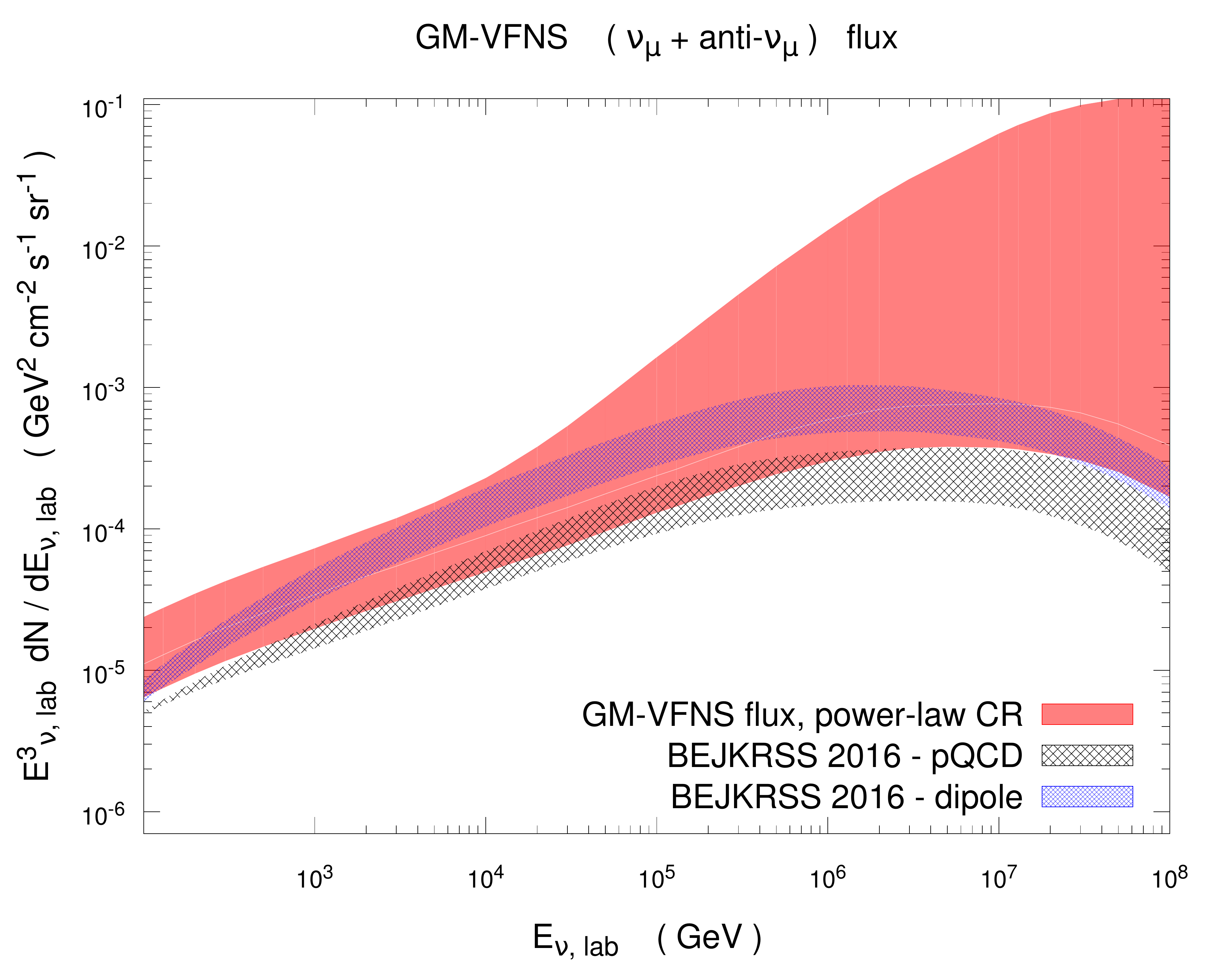}
\end{center}
\caption{\label{promptconven} Prompt-neutrino fluxes with their uncertainties according to the GM-VFNS approach 
compared with predictions for conventional neutrino flux and other predictions using the superposition model (left panel). Further comparisons with independent computations of prompt-neutrino fluxes~\cite{Bhattacharya:2016jce} using either nuclear PDFs (grey area) in an alternative GM-VFNS approach (FONLL~\cite{Cacciari:2012ny}) or three different dipole models (spanning the violet area) are shown in the right panel. See text for more details.}  
\end{figure}

In Fig.~\ref{promptconven} we show the prompt-($\nu_\mu$ + $\bar{\nu}_\mu$) flux as predicted in this contribution with its QCD uncertainties, in comparison with some other predictions: the GMS flux of Ref.~\cite{Garzelli:2015psa}, the PROSA flux of Ref.~\cite{Garzelli:2016xmx} and the BERSS flux of Ref.~\cite{Bhattacharya:2015jpa}.
It is interesting to observe that, notwithstanding the differences in shapes between the GM-VFNS predictions and those of Refs.~\cite{Garzelli:2015psa,Garzelli:2016xmx}, the transition energy, i.e.\ the energy where the prompt-neutrino flux becomes larger than the conventional one, evaluated starting from the Honda predictions of Ref.~\cite{Honda:2006qj} re-weighted to the H3a CR primary spectrum, is very similar in the three cases and amounts to $E_{\nu,\, trans}\approx 8 \cdot 10^5$~GeV. 

While the GM-VFNS predictions, as well as those of Refs.~\cite{Garzelli:2015psa, Garzelli:2016xmx,Bhattacharya:2015jpa}, are based on the superposition model, i.e.\ the cross section for $pA$ interactions is written as a superposition of $pp$ collisions, the predictions of Ref.~\cite{Bhattacharya:2016jce} are based on the use of nuclear PDFs. 
We compare the GM-VFNS predictions with those of Ref.~\cite{Bhattacharya:2016jce} in  the right panel of Fig.~\ref{promptconven}. The central predictions on the basis of nuclear PDFs are suppressed with respect to our central predictions 
for all considered energies. However, present uncertainties on nuclear PDFs are  definitely underestimated. The nuclear PDF fits, differently from the nucleon PDF fits, are still at the infancy of their development, due to both the lack of experimental data using nuclear targets and the uncertain theoretical description of nuclear matter effects as compared to the $ep$, $pp$ and $p\bar{p}$ cases.  We can thus still conclude that our predictions would be compatible with those based on nuclear PDFs, if one would take into account the uncertainties on the latter in a more reliable way. 
In Fig.~\ref{promptconven}, we also show predictions obtained by the dipole models considered in Ref.~\cite{Bhattacharya:2016jce}, 
which allow to evaluate cross-sections for heavy-quark production in an alternative way with respect to the standard partonic pQCD approach. 
These predictions, with their uncertainties, are fully included in the GM-VFNS uncertainty band for all values
$E_\nu \in [10^2$--$10^8]$.

\bibliographystyle{JHEP}
\bibliography{con}

\end{document}